\documentclass[twocolumn]{aastex62}

\usepackage{amsmath}
\newcommand\kmps{\mbox{km\,s${}^{-1}$}}

\shorttitle{Multi-epoch Optical Images of IRC+10216}
\shortauthors{KIM ET AL.}

\begin{document}
\title{Multi-epoch Optical Images of IRC+10216 Tell About the Central Star And the Adjacent Environment}

\author{Hyosun Kim}
\affiliation{Korea Astronomy and Space Science Institute, 776,
  Daedeokdae-ro, Yuseong-gu, Daejeon 34055, Republic of Korea}
\affiliation{Institute of Astronomy and Astrophysics, Academia Sinica,
  11F of Astronomy-Mathematics Building, AS/NTU,
  No.1, Sec. 4, Roosevelt Rd, Taipei 10617, Taiwan, R.O.C.}

\author{Ho-Gyu Lee}
\affiliation{Korea Astronomy and Space Science Institute, 776,
  Daedeokdae-ro, Yuseong-gu, Daejeon 34055, Republic of Korea}

\author{Youichi Ohyama}
\affiliation{Institute of Astronomy and Astrophysics, Academia Sinica,
  11F of Astronomy-Mathematics Building, AS/NTU,
  No.1, Sec. 4, Roosevelt Rd, Taipei 10617, Taiwan, R.O.C.}

\author{Ji Hoon Kim}
\affiliation{Subaru Telescope, National Astronomical Observatory of Japan, 650 North A'ohoku Place, Hilo, HI 96720, US}
\affiliation{METASPACE Inc., 36, Nonhyeon-ro, Gangnam-gu, Seoul, Republic Of Korea}

\author{Peter Scicluna}
\affiliation{Institute of Astronomy and Astrophysics, Academia Sinica,
  11F of Astronomy-Mathematics Building, AS/NTU,
  No.1, Sec. 4, Roosevelt Rd, Taipei 10617, Taiwan, R.O.C.}
\affiliation{European Southern Observatory, Alonso de Cordova 3107,
  Santiago RM, Chile}

\author{You-Hua Chu}
\affiliation{Institute of Astronomy and Astrophysics, Academia Sinica,
  11F of Astronomy-Mathematics Building, AS/NTU,
  No.1, Sec. 4, Roosevelt Rd, Taipei 10617, Taiwan, R.O.C.}

\author{Nicolas Mauron}
\affiliation{Laboratoire Univers et Particules, Universite de Montpellier and CNRS, Batiment 13, CC072, Place Bataillon, 34095 Montpellier, France}

\author{Toshiya Ueta}
\affiliation{Department of Physics and Astronomy, University of Denver,
  2112 E Wesley Ave., Denver, CO 80208, U.S.A.}

\correspondingauthor{Hyosun Kim}
\email{hkim@kasi.re.kr}

\begin{abstract}
  Six images of IRC+10216 taken by the {\it Hubble Space Telescope} at three epochs in 2001, 2011, and 2016 are compared in the rest frame of the central carbon star. An accurate astrometry has been achieved with the help of {\it Gaia} Data Release 2. The positions of the carbon star in the individual epochs are determined using its known proper motion, defining the rest frame of the star. In 2016, a local brightness peak with compact and red nature is detected at the stellar position. A comparison of the color maps between 2016 and 2011 epochs reveals that the reddest spot moved along with the star, suggesting a possibility of its being the dusty material surrounding the carbon star. Relatively red, ambient region is distributed in an $\Omega$ shape and well corresponds to the dusty disk previously suggested based on near-infrared polarization observations. In a larger scale, differential proper motion of multiple ring-like pattern in the rest frame of the star is used to derive the average expansion velocity of transverse wind components, resulting in $\sim$\,12.5\,km\,s$^{-1}$\,($d$/123\,pc), where $d$ is the distance to IRC+10216. Three dimensional geometry is implied from its comparison with the line-of-sight wind velocity determined from half-widths of submillimeter emission line profiles of abundant molecules. Uneven temporal variations in brightness for different searchlight beams and anisotropic distribution of extended halo are revisited in the context of the stellar light illumination through a porous envelope with postulated longer-term variations for a period of $\lesssim10$ years.
\end{abstract}

\keywords{circumstellar matter
  --- stars: AGB and post-AGB
  --- stars: individual (IRC+10216) 
  --- stars: late-type
  --- stars: mass-loss
  --- stars: winds, outflows
}

\section{INTRODUCTION}\label{sec:int}

IRC+10216 (or CW Leonis) is a stellar system that includes a post-main-sequence star and a thick dusty envelope surrounding the star. The circumstellar envelope (CSE) is the material expelled from the star with an increasing mass loss rate that reaches its extreme when the star is at the asymptotic giant branch (AGB) of the stellar evolutionary track.
Since IRC+10216 is the nearest carbon-rich AGB star (also known as carbon star) at a distance of 123\,$\pm$\,14\,pc \citep{gro12}, this object has been used as an important laboratory for various experiments. Multi-epoch multi-wavelength observations on IRC+10216 have been carried out to study its circumstellar dust, various molecules and chemical elements \citep[e.g.,][]{gue77,jur83,gue11}, as well as the geometry of circumstellar medium relevant to the light pulsations, partial shells, and wind kinematics \citep[e.g.,][]{cra87,mau99,dec11,cer15}.

IRC+10216 revealed its extended CSE brightened by dust-scattered, ambient Galactic light in deep optical-band observations \citep[e.g.,][]{mau99}. The CSE is detected out to about 200\arcsec\ (or 0.1\,pc at a distance of 123\,pc) with a roughly spherical shape, reflecting a history of outflow at a speed of $\sim14\,\kmps$ over the past $\sim8,000$\,yr. A series of incomplete thin shells are well detected over the entire extent of the optical envelope with shell separations of $\sim5\arcsec$--20\arcsec, corresponding to $\sim200$--800\,yr.
These shells are also studied through (sub-)millimeter line observations with the advantage of ubiquity of CO molecule \citep[e.g.,][]{fon03,gue18}. The shells outside 40\arcsec\ up to $\sim3\arcmin$, verified by the Submillimeter Array, are nearly concentric and regularly spaced at an average separation of 16\arcsec\ (timescale $\sim$\,650\,yr). Closer to the star ($<40\arcsec$), the incomplete and less-regular shells (or arcs), traced by the Atacama Large Millimeter/submillimeter Array at an angular resolution of 0\farcs3, are more closely packed with separations decreasing down to only $\sim2\arcsec$ (timescale $\sim$\,80\,yr) within 10\arcsec\ from the carbon star.
The hypothesis of a central binary system has been strongly bolstered by these optical and radiomillimeter observations, and the findings of shells around other AGB stars for which the binarity is also proposed: AFGL 3068 \citep{kim12b,kim17}, R Scl \citep{mae12}, CIT 6 \citep{kim13,kim15b}, EP Aqr \citep{hom18}, and GX Mon \citep{ran20}. It is supported by hydrodynamical models that predict nested spiral-like shells \citep[e.g.,][]{mas99,edg08,kim12a,kim19}. The orbital period of 55 yr at present, and 800 yr in the past, is suggested to explain the timescales of separations between shells \citep{cer15,dec15}.

The central part of IRC+10216 has been subject to studies {\it via} observations at all available wavelengths.
Among them, in all of the optical to near-infrared observations, the carbon star was completely obscured and a sub-arcsecond bipolar-like structure appeared \citep[see e.g.,][]{han98,ski98,mau00,lea06}. However, the images taken later in various near-infrared monitoring programs revealed that the near-infrared brightened regions are clumpy and evolving fast. In particular, the relative brightnesses of bright clumps vary without leaving any persistent features \citep{ste16}.
A drastic change of optically bright features in the central 1\arcsec\ core region from 1998--2001 to 2011 was also identified \citep{kim15a}. All available light curves of IRC+10216 in the optical and infrared were also revisited in this paper. One of their interesting findings is the brightening trends in the periods of 1985--1989 ($-0.3$ to $-0.1$\,mag yr$^{-1}$ in $JHK$ bands) and 2005--2013 ($-0.16$\,mag yr$^{-1}$ in the optical) with, in contrast, no brightening in 1999--2008 in all $JHKLM$ light curves. This gradual brightening in a timescale of $\sim10$\,yr was interpreted as a progressive rarefaction of the inner envelope along the line of sight toward the stellar position.

The substantial changes of the inner core images, in addition to the longer term variations in the light curves and the shell intervals, inspired us to pursue a {\it Hubble Space Telescope} ({\it HST}) monitoring program.
In this paper, we report new {\it HST} imaging observations taken in 2016 with three filters in the optical to near-infrared wavelengths. These images are compared with the corresponding band images taken previously at the epochs of 2011 and 2001. The comparison of these epochs with the intervals of 10 and 5 years gives an additional hint for the presence of a cycle in brightness with a decadal timescale.
In Section\,\ref{sec:obs}, the details of the observations in 2016 are described, along with a brief summary of the observations in 2001 and 2011 (Section\,\ref{sec:set}). The astrometry method is explained in Section\,\ref{sec:ast}, and the units of energy flux is converted to Jansky (Section\,\ref{sec:flx}). Findings from inspecting the images in small to large scales are demonstrated in Section\,\ref{sec:res}. A compact, red spot is detected at the expected stellar position in the 2016 epoch (Section\,\ref{sec:agb}). Color maps in different epochs are examined in regard to the dusty disk/torus surrounding the star (Section\,\ref{sec:dsk}). In Section\,\ref{sec:exp}, the recurrent ring-like pattern imprinted in the CSE is used to determine the expansion velocity of wind components moving in the midplane. This transverse wind velocity is compared to the line-of-sight wind velocity derived from molecular line observations, and the implication is discussed. Over larger scales, several searchlight beams are identified in Section\,\ref{sec:bem}, including two beams previously appearing aligned in a bipolar shape and giving a misleading impression. Rapid changes of overall halo brightness distribution within 5 years, significantly shorter than the dynamic timescale, are detected and discussed in Section\,\ref{sec:cse}. Finally, we summarize the main results in Section\,\ref{sec:sum}.

\section{OBSERVATIONS AND DATA REDUCTION}\label{sec:obs}
\subsection{Observations}\label{sec:set}

Three {\it HST} images were obtained on 2016 May 17 (epoch 2016.38) with the Wide Field Camera 3 (WFC3) and in three filters: F606W, F814W, and F098M. Their central wavelengths (and widths) are 5841 (2345), 8236 (2543), and 9847 (1693) \AA, respectively.
To avoid saturation at the core, we observed near a minimum in the stellar light curve, at a phase of 0.12, according to an ephemeris based on the Catalina Sky Survey \citep[see Appendix\,\ref{sec:lcr} for more details]{dra14}.
For F606W and F814W, the individual exposure times were 350 seconds with the 2-point WFC3-UVIS-GAP-LINE dithering pattern, giving a total exposure of 700 seconds for each filter. For F814W, a post-flash of 4 electrons was applied to mitigate the charge transfer efficiency loss in the detector. For F098M, the minimum exposure time 2.932 seconds in rapid mode was adopted with 6 iterations per dithering position of the 2-point WFC3-IR-DITHER-BLOB pattern, therefore the total exposure time was 35.187 seconds.
The pixel size is 0\farcs04 for F606W and F814W, and 0\farcs13 for F098M. The field sizes are $150\arcsec\times150\arcsec$ for all filters.
Our new images are compared to those of previous observations retrieved from the Mikulski Archive for Space Telescopes. Details are given in Table\,\ref{tab:list}.
We obtained the calibrated, distortion-corrected archival F606W and F814W images processed with the standard STScI calwf3 v3.1.6 pipeline \citep{gen18}.

\subsection{Astrometry}\label{sec:ast}

We aligned the six {\it HST} images by calculating offsets between images using the TWEAKREG task in the DRIZZLEPAC software package \citep{gon12}.
We first considered the positions of field stars from the {\it Gaia} Data Release 2 \citep{gai16,gai18} and took into account their proper motions to obtain their 2016 coordinates. Ten field stars were used to register the world coordinate system (WCS) of the 2016 F606W image, and we take that image as our reference image in the following. The 2016 F814W and F098M images could then be placed in the same frame by using twelve and six field stars, respectively. Concerning the 2001 and 2011 images, different observational depths and sky coverages restrict the number of field stars that can be used for the image alignment. Thus, we instead used the central positions of a few field galaxies with respect to the 2016 F606W image. Finally, fifteen stars were used for alignment between 2011 images with different filters.

After registering the WCS for all images, we calculated the expected WCS position of the carbon star at each epoch (2001.02, 2011.42, and 2016.38). For this, we applied the proper motion of the star $\rm (\mu_{R.A.},\,\mu_{Dec.})$ of $(+35\pm1,\,+12\pm1)$\,mas\,yr$^{-1}$ (p.m., hereafter), derived using accumulated radio continuum positions, to the position of the radio continuum peak accurately determined with the Very Large Array (VLA) at the epoch 2006.16 \citep{men12}. The p.m.-corrected VLA position defines the ``stellar'' position used below and all following figures are centered on this position at the 2016.38 epoch, unless noted otherwise.

The coordinate shifts for image alignments were calculated in pixel scales as archived. The positions of field stars in our astrometry have an uncertainty of $\sim0\farcs02$ compared to the positions of those stars in {\it Gaia} catalog. Similar amount of uncertainty exists in the image alignments between different epochs using the positions of stars or galaxies in the {\it HST} images. Taking into account the uncertainties in the alignment between {\it HST} images, the {\it Gaia} astrometry, the measurement of radio continuum peak position, and the p.m.\ derivation, the error of stellar position in our images is not more than one pixel.

\subsection{Flux calibration}\label{sec:flx}

The image brightness values are converted from counts per second to energy flux unit, Jansky, based on the PHOTFLAM and PHOTPLAM values stored in the image header.
All images are resampled to a $0\farcs04\times0\farcs04$ pixel using the \texttt{reproject} package in the Python code, and the flux per pixel is corrected accordingly.
The sky background, estimated from the peak in the flux histogram, is additionally subtracted. The subtracted sky value is in the level of $7\sigma$ for the 2001 image but is negligible for the other five images. The noise level $\sigma$ is estimated for each image; all of them are in the order of $10^{-9}$\,Jy/pixel.

\section{RESULTS AND DISCUSSION}\label{sec:res}
\subsection{The position of the central star}\label{sec:agb}

\begin{figure*} 
  \epsscale{1.19}
  \plotone{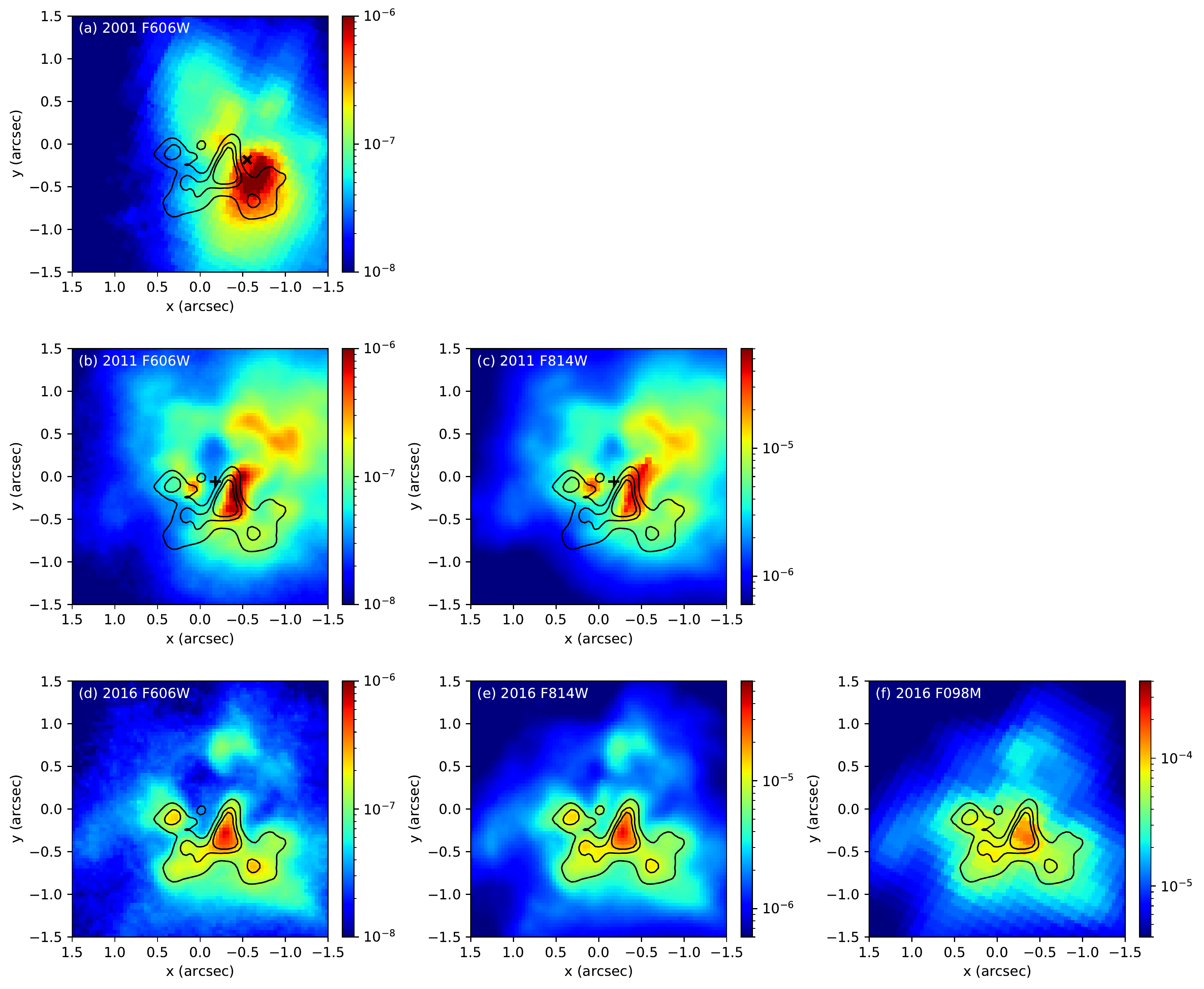}
  \caption{\label{fig:all}%
    Central field of size $3\arcsec\times3\arcsec$: (a) in 2001 with F606W
    filter, (b) and (c) in 2011 with F606W and F814W filters, and (d), (e)
    and (f) in 2016 with F606W, F814W and F098M filters. In all panels, the
    pixel size is $0\farcs04\times0\farcs04$. North is up and east to the left.
    The ranges of color bars in logarithmic scales are from $10^{-8}$ to
    $10^{-6}$\,Jy/pixel for the F606W images displayed in (a), (b) and (d),
    $6\times10^{-7}$ to $6\times10^{-5}$\,Jy/pixel for the F814W images in
    (c) and (e), and $4\times10^{-6}$ to $4\times10^{-4}$\,Jy/pixel for
    the F098M image in (f).
    The black contours are isophotes of the 2016 F814W image in panel (e)
    at levels of $6\times$, $11\times$, and $16\times10^{-6}$\,Jy/pixel.
    The same contours are overplotted in all panels for comparison. The
    coordinate origin in all panels is at the predicted stellar position
    in 2016. The predicted stellar positions in the individual epochs are
    marked by a cross ($\times$) symbol in panel (a) and a plus (+) symbol
    in panels (b) and (c). This figure shows the drastic evolution of the
    central region in timescale of years.
  }
\end{figure*}

Figure\,\ref{fig:all} exhibits the central region of IRC+10216 in the {\it HST} images taken (a) in 2001 with the F606W filter, (b) and (c) in 2011 with the F606W and F814W filters, respectively, and (d), (e), and (f) in 2016 with the F606W, F814W, and F098M filters, respectively.
The coordinate origins are all set to be the stellar position at the latest epoch 2016.38 and the $x$-axis and $y$-axis are along the right ascension (R.A.) and declination (Dec.), respectively. The cross ($\times$) symbol in Figure\,\ref{fig:all}(a) and plus (+) symbol in (b)--(c) indicate the stellar positions at the corresponding epochs that the individual images were taken, which are 2001.02 and 2011.42, respectively.

In 2001, the p.m.-corrected position of the star is to the north of the brightness peak in the optical wavelength.
The p.m.-corrected stellar position in 2011 is located between the optical brightness peaks and close to the local minimum in brightness. In the 2016 epoch, similar to the 2011 epoch, the stellar position is to the east of the optical brightness maximum. One interesting point in 2016 is the coincidence of the stellar position with a local peak in the F814W image, which is noticeable in the contours of 2016 F814W image overlaid in all panels in Figure\,\ref{fig:all}. 

\begin{figure*} 
  \plotone{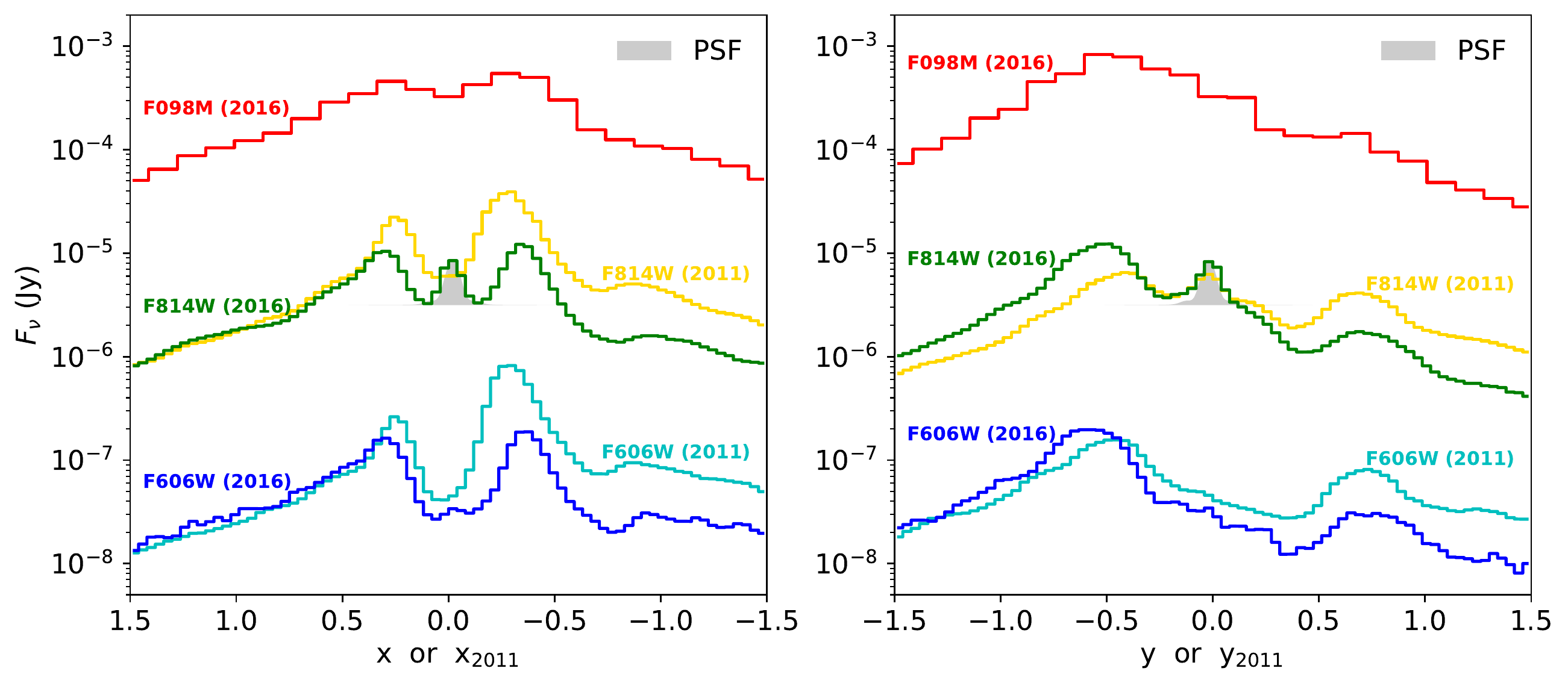}
  \caption{\label{fig:prf}%
    Brightness profiles of central region of IRC+10216.
    The profiles of F606W (blue), F814W (green), and F098M (red) images
    taken in the 2016 epoch are displayed along $x$- (left panel) and
    $y$-axes (right panel). A small brightness bump at $(x,y)=(0,0)$ is
    identified in the F814W profiles (green), whose size is comparable
    to, or slightly larger than, the PSF size of detector (gray).
    For comparison, the corresponding profiles of images taken in the 2011
    epoch with the F606W (cyan) and F814W (yellow) filters are displayed,
    but in the coordinates ($x_{2011}$, $y_{2011}$) with the origin set at
    the stellar position at its own epoch.
    Plots are in the original pixel scales before regridding images.
  }
\end{figure*}

The brightness profiles of F606W, F814W, and F098M images taken in 2016 are plotted in Figure\,\ref{fig:prf} by blue, green, and red curves, respectively, along $x$-axis ({\em left} panel) and $y$-axis ({\em right} panel). A bump at the stellar position is clearly seen in the F814W profiles (green curves) but not obvious in the F606W profile. The non-detection in the F098M profile is inconclusive because of the lower resolution of the image at IR wavelengths. The compactness of this local brightness peak is apparent in Figure\,\ref{fig:prf}. Its full-width at half-maximum (FWHM) above the adjacent diffuse emission is $\la0\farcs10$, which is slightly larger than the FWHM of standard point spread function (PSF) model, $\sim$\,0\farcs08.\footnote{We use the standard PSF downloaded from https://stsci.edu/hst/instrumentation/wfc3/data-analysis/psf estimated near the detector position of our target source. Because the diffuse emission is anisotropic about the stellar position of IRC+10216, we measure the FWHMs of brightness profiles in several directions with different definitions for the levels of diffuse emission, resulting in 0\farcs089--0\farcs096. In another way, the azimuthally-averaged brightness profile is used and its first local minimum is defined as the level of diffuse emission, resulting in the FWHM of 0\farcs084.} For comparison, this size is slightly larger than the size of radio continuum \citep[0\farcs083;][]{men12} and smaller than the inner size of dust formation region \citep[$\sim$\,0\farcs2;][]{dec15}.

In the same figure, the brightness profiles of 2011 images with F606W and F814W filters in cyan and yellow, respectively, are overlaid but along the $x_{2011}$- and $y_{2011}$-axes which represent the coordinates with the origin at the stellar position at the 2011.42 epoch. In the right panel, i.e., along the $y_{2011}$-axis, a small bump is seen at the p.m.-corrected stellar position, which might imply that the flux at the stellar position has increased since 2011 and has moved along with the central star at the same p.m.\ speed. 

\begin{figure*} 
  \epsscale{1.1}
  \plotone{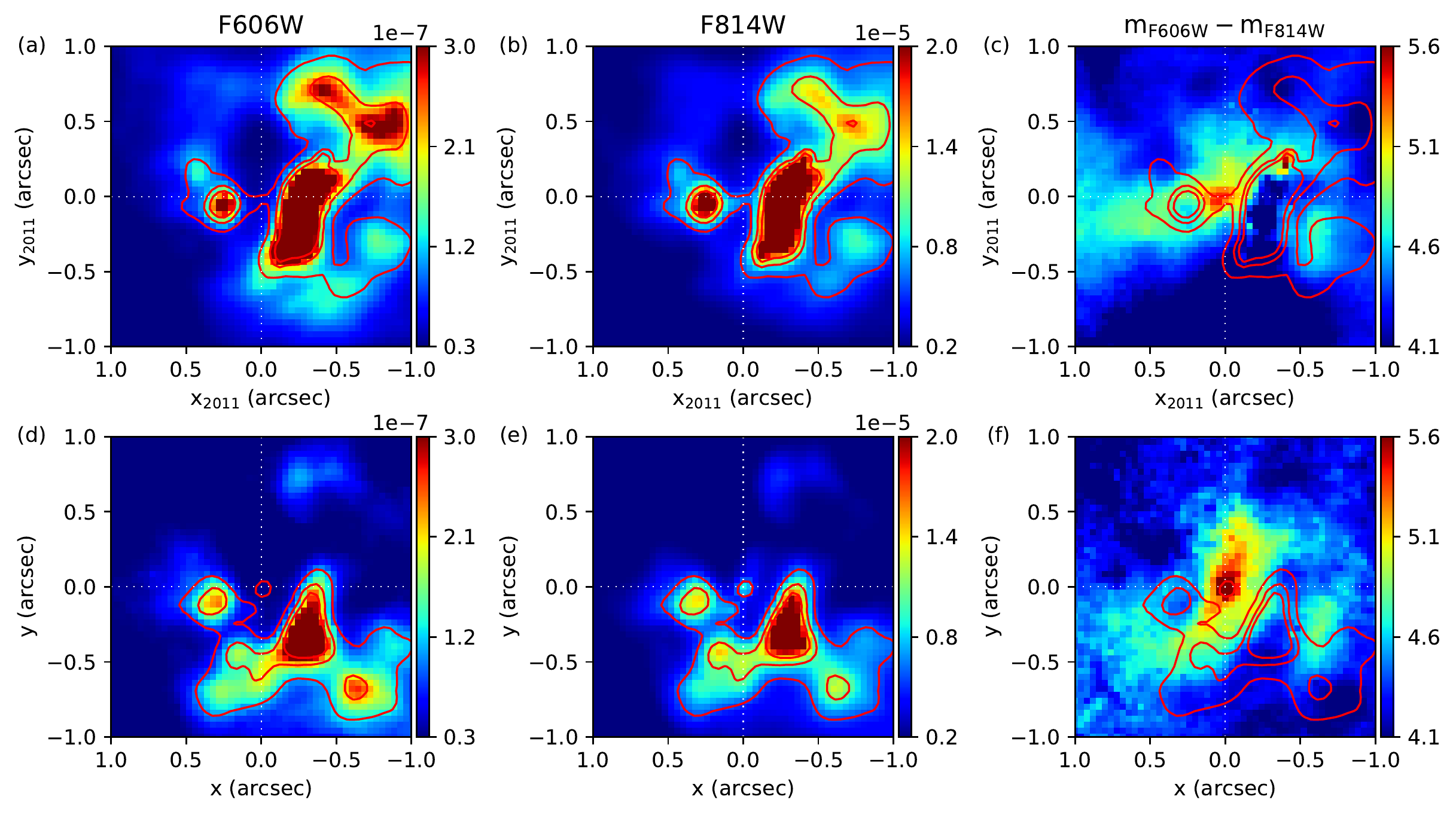}
  \caption{\label{fig:clr}%
    Optical images and their color maps: (a)--(c) in 2011, and
    (d)--(f) in 2016. Panels (a)--(b) and (d)--(e) are the same
    as Figure\,\ref{fig:all}(b)--(c) and Figure\,\ref{fig:all}(d)--(e),
    respectively, but in linear color scales ranging in (a) and (d)
    from $3\times10^{-8}$ to $3\times10^{-7}$\,Jy/pixel
    and in (b) and (e)
    from $2\times10^{-6}$ to $2\times10^{-5}$\,Jy/pixel.
    Color maps in panels (c) and (f) are defined as the log ratios of
    intensities F814W/F606W in 2011 and 2016, respectively, exhibiting
    reddened property at the stellar positions relative to the
    neighboring material. For reference, the contours of F814W
    intensity maps are superimposed in panels (a)--(c) and
    (d)--(f), respectively, at levels of $6\times$, $11\times$, and
    $16\times10^{-6}$\,Jy/pixel. The coordinate origins are at
    the predicted stellar positions in the corresponding epochs.
  }
\end{figure*}

In Figure\,\ref{fig:clr}, we exhibit the F606W and F814W images (same as in Figure\,\ref{fig:all} but in linear color scales) and the brightness ratio map between these images (i.e. color maps) in the 2011 and 2016 epochs.
The brightness peak at the stellar position in the 2016 epoch is well identified in the F814W image in linear color scale (see Figure\,\ref{fig:clr}(e)).
In the F606W--F814W color map of 2016 epoch, Figure\,\ref{fig:clr}(f), this central brightness peak turns out to be redder than the neighboring material. Interestingly, the reddest spot in 2011 was also near the p.m.-corrected stellar position at that epoch (see Figure\,\ref{fig:clr}(c)). One possible interpretation is that the distinct brightness peak with very red color has moved along with the carbon star from 2011 to 2016. We postulate that this reddest, local brightness peak is the very close circumstellar material attached to the carbon star.

\subsection{Dusty disk/torus}\label{sec:dsk}

In both 2011 and 2016 epochs, the F606W--F814W color maps exhibit not only the reddest point located at the p.m.-corrected stellar positions but also some extension of the reddened regions in northwest-to-southeast direction (see Figure\,\ref{fig:clr}(c) and (f)). In each epoch, the color map presents the elongated (banana-shaped) structure in a rather point-symmetric sense about the corresponding stellar position and also some extent to the west side of the optically brightest region, composing overall an Omega ($\Omega$) shape.

\begin{figure*} 
  \epsscale{1.17}
  \plotone{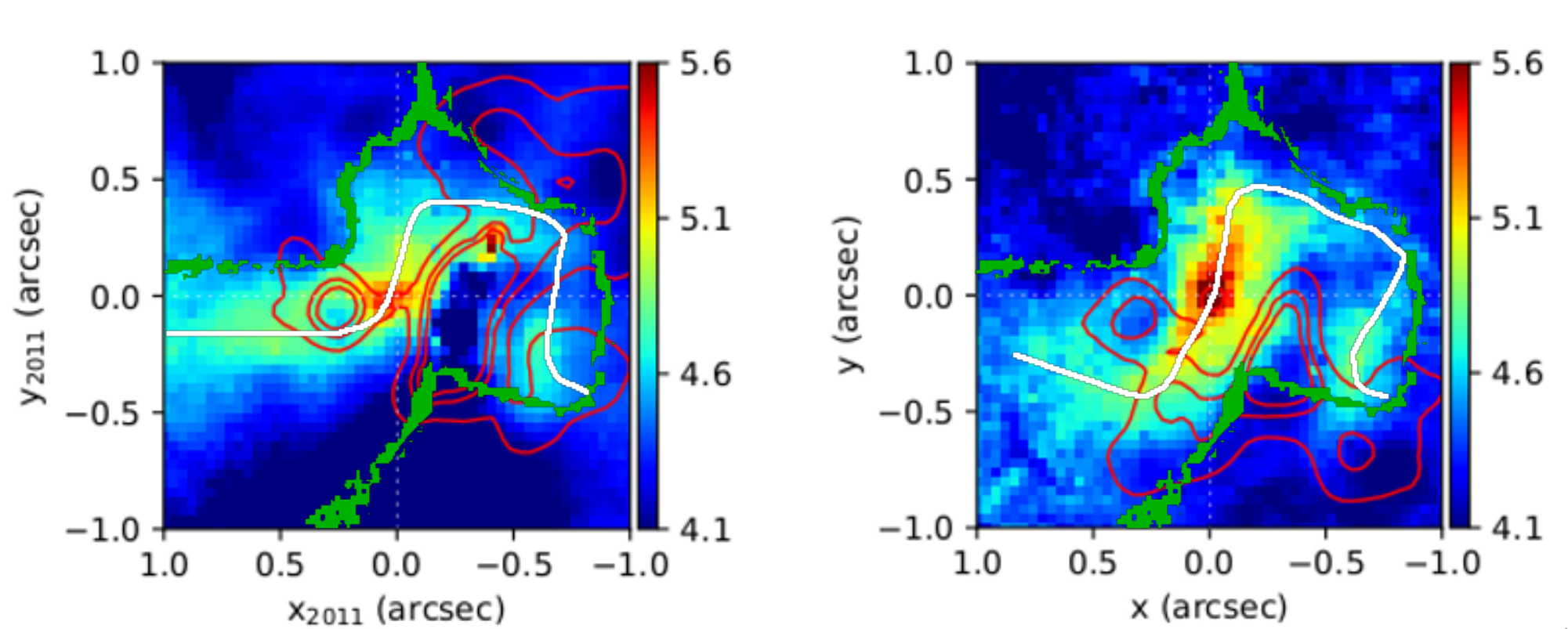}
  \caption{\label{fig:dsk}%
    Comparison of color maps from the {\it HST} images in ({\em left})
    2011 and ({\em right}) 2016. See the caption for Figure\,\ref{fig:clr}
    for more details. White solid lines in $\Omega$ shape trace the redder
    regions. Green patched line indicates the outer boundary of the
    dusty disk suggested by \citet{mur05} based on their polarization map in
    $H$-band of the Subaru telescope at the degree of the linear polarization
    of $\sim10\%$ (observation epoch of 2003). The maps in three epochs
    are registered in the rest frame of the carbon star by p.m.\ correction.
  }
\end{figure*}

In Figure\,\ref{fig:dsk}, we show that the relatively red regions in these two different epochs are located well within the dusty disk/torus suggested by \citet{mur05} based on the degree of linear polarization in the epoch of 2003. The structures in these three epochs (2003, 2011, and 2016) are consistent with one another, except for the optically brightest regions in 2011 and 2016, which are relatively bluer probably due to the significant dust scattering process on the circumstellar dust clouds in front of midplane structures.

\subsection{Wind expansion}\label{sec:exp}

\subsubsection{Background}

It is important to understand the three-dimensional morphology of the CSE of AGB stars, in particular in the framework of binary population in evolved phases of stars.
In theory for a spiral-shell pattern induced by a binary system, the expansion velocity of the pattern near the orbital plane is highly influenced by the binary orbital velocity but the fluid velocity near the orbital axis is not much influenced \citep[e.g.,][]{sok94,mas99,kim12a,kim19}, possibly compressing the CSE toward the orbital plane.
There was a study on the flattening of the envelope based on deep optical images of 22 AGB stars by measuring the shape ellipticity defined as the ratio between major and minor axes of iso-intensity contours \citep{mau13}. About 20\%\ of their target AGB stars show images with ellipticities greater than 1.2, which is suggested to be a lower limit of the true occurence of ellipticity because of the line of sight geometry. The authors interpret this percentage as the population of binaries whose envelopes are flattened by a companion.

The three-dimensional morphology of the CSE of an AGB star can be inferred at the first order from measuring the expansion velocities of transverse and radial (line-of-sight) components of the circumstellar wind flows. The line-of-sight velocity is easily estimated from the spectral width of observed molecular line emission \citep[e.g.,][]{kna82,lou93}. For the transverse velocity measurement, the differential proper motions of circumstellar features lying on the plane of the sky can be used. In particular, the ring-like pattern, if exists, is extremely useful. The ring-like pattern in an optical image traces the region brightened by scattering process with an efficiency depending on the dust column density. It represents the tangential section of the shell-like pattern in three-dimensional space onto the midplane. Therefore, the differential location of the ring-like pattern in two different epochs can be used to measure the velocity of the flow located near the midplane. 

For example, \citet{bal12} and \citet{uet13} measured the differential proper motions of circumstellar arcs of a preplanetary nebula, the Cygnus Egg Nebula, using {\it HST} images from two epochs. The former treated the whole ensemble of ring-like pattern and concluded that the transverse velocity is 0\farcs07 per 6.65-yr from 2002 to 2009, corresponding to $\sim$\,20\,\kmps\ if the distance of 420\,pc is assumed. On the other hand, the latter treated various arc segments along poles in the images taken in 1995 and 2002 and found the constancy of their velocities along radius, as $\sim$\,10\,\kmps\ assuming a distance of 420\,pc. More recently, \citet{gue20} made a similar analysis for one AGB CSE and three planetary nebulae with several arc segments per target and the derived expansion velocities of these segments showed large variations along position angle. The radius dependencies (at the given position angles) were insignificant, similarly to the trend in \citet{uet13}.

A similar method is applied to the multiple ring-like pattern appearing in the CSE of IRC+10216. We first measure the expansion velocity averaged over the plane of the sky, and then compare the velocity measurements in 36 individual sectors, each of which has an opening angle of 10\arcdeg.
For this purpose, we have employed the F814W images at the 2011 and 2016 epochs, whose observation times were sufficient to detect the light from the ring-like patterns in both epochs. The observation performed in 2016 is shallower and the following analysis is made within $r\sim5\arcsec$, where the F814W brightness reaches $\sim7\sigma$ detection.

\subsubsection{Average expansion velocity}\label{sec:vex}

\begin{figure*} 
  \epsscale{1.17}
  \plotone{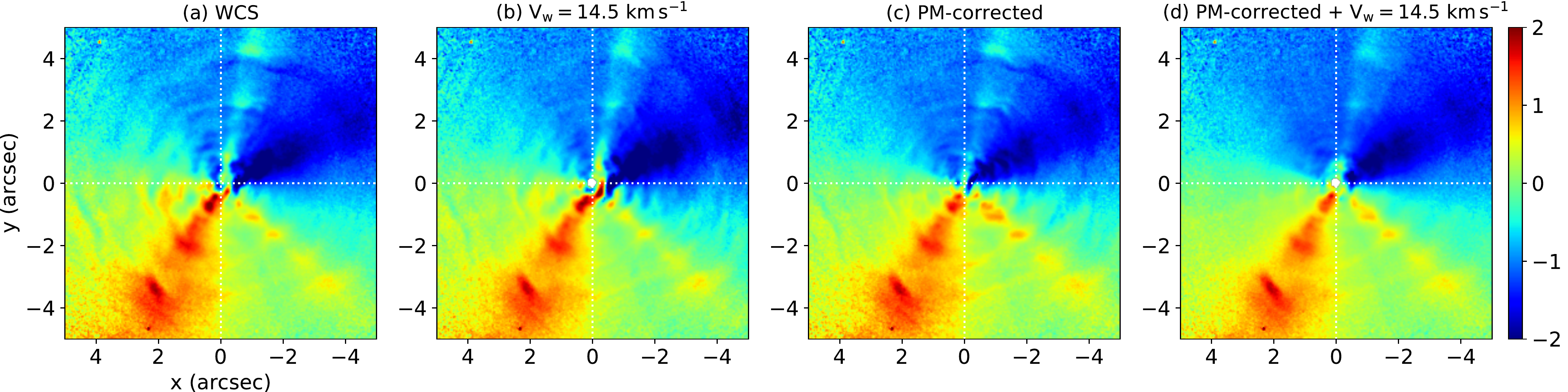}
  \caption{\label{fig:cmp}%
    Magnitude differences between F814W images taken in 2011 and in 2016, 
    $\rm m_{2011}-m_{2016}$, with four different alignments: 
    the coordinates of images in the two different epochs are aligned
    (a) in WCS, and after applying to the 2011-epoch image (b) a wind
    expansion of 14.5\,\kmps\ or (c) the stellar p.m.\ correction, and
    (d) both of them. A larger value indicates that the corresponding
    pixel is brighter in the 2016 image relative to that in the 2011
    image. The pixels displayed in white color indicate the central
    region of which the information before expanding is lacking in the
    2011 image. Image size is $10\arcsec\times10\arcsec$. 
  }
\end{figure*}

In Figure\,\ref{fig:cmp}, we present the residuals of the 2011's image magnitudes\footnote{Magnitude m is defined as ${\rm m}=-2.5\log_{10}F$, where $F$ indicates the image value in units of Jy/pixel.} in F814W after subtracting the 2016's magnitudes in the same band, $\rm m_{2011}-m_{2016}$, with four different alignments between the images at the two epochs.
Figure\,\ref{fig:cmp}(a) is the epoch-difference map in the WCS, as registered in Section\,\ref{sec:ast} using the positions of field stars and galaxies, in which the positions of the carbon star are offset between the two epochs.
For Figure\,\ref{fig:cmp}(b), before subtracting the 2016's magnitude image, the 2011's image is enlarged by moving the image grids in the radial direction by the radius $\delta r$, given by the multiplication between the assumed expansion velocity of the wind, $V_w$, and the time difference between 2011 and 2016 epochs, $\delta t = 4.95$ yr:
\begin{equation}
  \left( \frac{\delta r}{\rm arcsec} \right)
  = 0.12
    \left( \frac{V_w}{14\,\rm km/s} \right)
    \left( \frac{\delta t}{4.95\,\rm yr} \right)
    \left( \frac{d}{123\,\rm pc} \right)^{-1},
\end{equation}
where $d$ is the distance of IRC+10216. The image is regridded as the CSE expands, therefore the flux per pixel is scaled by a factor inversely proportional to the square of radius.
For Figure\,\ref{fig:cmp}(c), instead of applying the expansion of the pattern, the coordinates of the 2011's image are linearly transferred to ($x_{2011}$, $y_{2011}$)-coordinates, which superimposes the stellar positions between the two epochs based on the p.m.\ during the 4.95 years.
Finally, Figure\,\ref{fig:cmp}(d) shows the epoch-difference map after correcting both the p.m.\ of the star and the expansion of the ring-like pattern.
In Figure\,\ref{fig:cmp}(d), it is seen that most of the ring-like pattern is clearly cancelled out with both the p.m.\ correction \citep[based on][]{men12} and the wind expansion with a constant velocity of 14.5\,\kmps, albeit measured for the line of sight velocity \citep[see e.g.,][]{dec15}. In contrast, the corresponding epoch-difference maps either by excluding the p.m.\ correction or by setting a negligible expansion velocity always severely leave the ring-like patterns (Figure\,\ref{fig:cmp}(a)--(c)).

Note that even in Figure\,\ref{fig:cmp}(d), some conical regions (not in a shape of rings) have remained; the northwest, southeast, and southwest beams are particularly noticeable. These features exist in both 2011 and 2016 images but at different brightness (dubbed ``searchlight beams''; see Section\,\ref{sec:bem} for details). Here, in this section, in verifying how well the ring-like pattern is cleared up, these conical features act just as contamination. In addition to the conical structures, it is also noticeable that the remaining brightness distribution in Figure\,\ref{fig:cmp}(d) has a systematic gradation as a function of position angle (see Section\,\ref{sec:cse}). These extended brightness variations are additionally removed for the analysis below.

In order to estimate the expansion velocity of the wind in the midplane, we use the ring-like patterns revealed in the F814W images in 2011 and 2016 that are regridded with the coordinate origins at the p.m.-corrected stellar positions (see Section\,\ref{sec:ast}). In Figure\,\ref{fig:rtd}(a)--(b), these images are transformed to the polar coordinates, i.e., radius versus position angle (PA)\footnote{PA is measured from north to the east.}. And the 2011 image in the polar coordinates is shifted along radius by 4.95-yr times the expansion velocity as a free parameter. The pixel flux is scaled regarding the expansion. The magnitude difference between these two images is exhibited in Figure\,\ref{fig:rtd}(c). Figure\,\ref{fig:rtd}(d) is obtained by additionally subtracting the background gradation, defined as the blurred image with a Gaussian kernel along radius with its standard deviation of 0\farcs16. In Figure\,\ref{fig:rtd}(d), we still see some near-horizontal structures, corresponding to the remaining ring-like pattern, in the inner region ($r\la3\arcsec$). Above $r\sim3\arcsec$, random noise is dominant except for some remnants in the southern regions (PA of $\sim$ 140\arcdeg--250\arcdeg). In the remaining of this section, we try to find the expansion velocity minimizing the remnant pattern.

Because the standard deviation of brightness distribution in the residual image is dominated by random noise in some regions, it would not be the best indicator for how well the ring-like pattern is eliminated. We instead define the {\it remaining arcs} as the regions with the remnant brightnesses exceeding the predefined cutoff profile as a function of radius, and find the expansion velocity at which the excess emission within these {\it remaining arcs} is minimized. We repeat this procedure by changing some subsidiary parameters such as the extent of the image to be considered up to 10\arcsec\ and the standard deviation of Gaussian kernel to define the background gradation as twice larger. As the quantity to evaluate the minimization of the remnant pattern, we examine not only the summation of excess emission in the {\it remaining arcs} but also the summation of total emission in the {\it remaining arcs}, the standard deviation of the {\it remaining arcs}, and the standard deviation of the entire image. Considering the standard deviation of the six measurements as the uncertainty, the resulting transverse velocity is 12.5\,$\pm$\,0.3\,\kmps\ at a distance of 123 pc. Taking into account the distance uncertainty of $\pm$\,14\,pc in \citet{gro12}, we conclude the measured expansion velocity of the transverse wind components as 12.5\,$\pm$\,1.5\,\kmps.

It is slightly smaller than the expansion velocity along the line of sight, $\sim14.5\,\kmps$, derived from radio molecular line widths \citep[e.g.,][]{dec15}. This distinction between the velocities of radial (line-of-sight) and transverse wind components may indicate the non-spherical geometry of the overall CSE of IRC+10216. Moreover, the fact that the line-of-sight wind component is faster than the transverse wind component, likely implies that the orbital plane of the postulated binary stars is rather perpendicular to the plane of the sky. This is consistent with the model inclination proposed by \citet{dec15} based on the shape of position-velocity diagram of molecular line emission.
Alternatively, to construct a spherically symmetrical morphology and the corresponding isotropic expansion velocity, the distance of 143\,pc is required, which is beyond the uncertainty range of 123\,$\pm$\,14\,pc estimated by \citet{gro12}.

\subsubsection{Angle dependence of expansion velocity}

In Section\,\ref{sec:vex}, the expansion velocity is assumed to be independent of PA, which would be strictly true only for the case that the orbit is seen pole-on and the orbital shape is circular. For an eccentric-orbit binary, the directional dependency of wind expansion velocity can be complicated along the inclination angle, the location of the line of nodes of the orbit, and the location of the pericenter (Kim, {\it in prep.}). The discussion on the general trends of models through a parameter study is beyond the scope of this paper, but the velocity distributions in the models of \citet{kim19} seem to suggest that the transverse wind may have a minimum velocity toward the projected location of the pericenter of the carbon star orbit. Here we show as a case study the PA dependence of the transverse velocity of IRC+10216.

\begin{figure*} 
  \plotone{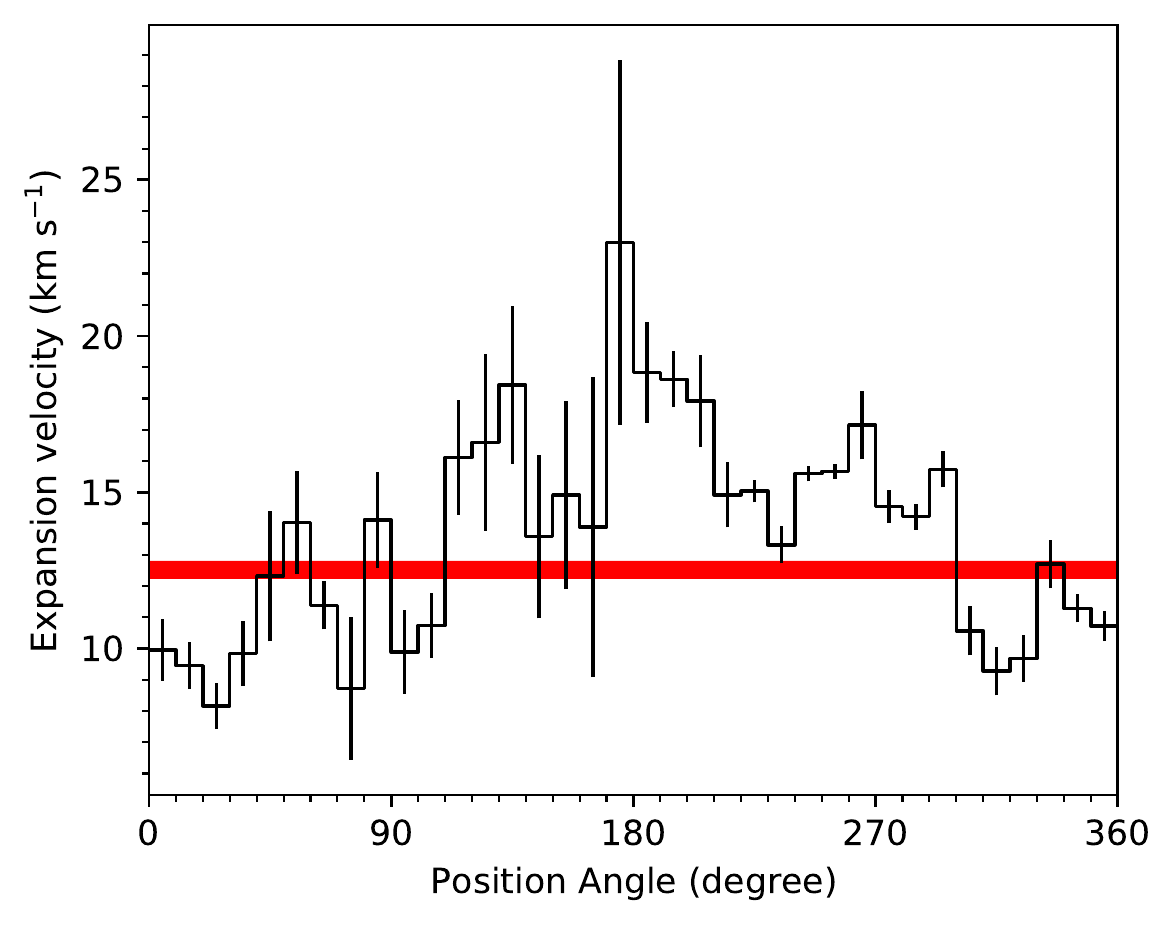}
  \caption{\label{fig:vex}%
    Expansion velocity derived from differential proper motion of ring-like
    pattern of IRC+10216. Changes of some subsidiary parameters are examined
    to define the mean and standard deviation of the resulting expansion
    velocities as denoted by the bar chart and error bars, respectively,
    as a function of position angle (black).
    An independent estimate for the average expansion velocity over the
    entire angles results in 12.5\,$\pm$\,0.3\,\kmps\ (red horizontal band).
  }
\end{figure*}

We repeat the same analysis as in the previous section but for 36 individual sectors with the opening angles of 10\arcdeg. The result is displayed in Figure\,\ref{fig:vex} including the vertical bars representing the standard deviation of six measurements by changing subsidiary parameters besides the expansion velocity. We note that the variation along the PA is about 7\,\kmps, which is comparable to the large PA dependence of 6--16\,\kmps\ found in the work by \citet{gue20} for NGC 7027, NGC 6543, and AFGL 3068. The overall trend in Figure\,\ref{fig:vex} is that the northern part is systemically lower than the southern part in the derived expansion velocity. If IRC+10216 were governed by a binary star system with the orbital shape in a circle, the wind expansion velocity would be maximized along the equatorial plane \citep[PA $\sim$\,110\arcdeg\ and $\sim$\,290\arcdeg, according to][]{dec15}, which is not the case seen in Figure\,\ref{fig:vex}. We speculate that it is an eccentric-orbit binary system with the projected position of the pericenter of the carbon star towards the northern direction. A concurrent data analysis for multi-epoch spectroscopic observations at high resolutions may be required for further discussion.

\subsection{Searchlight beams}\label{sec:bem}

\begin{figure*} 
  \epsscale{1.15}
  \plotone{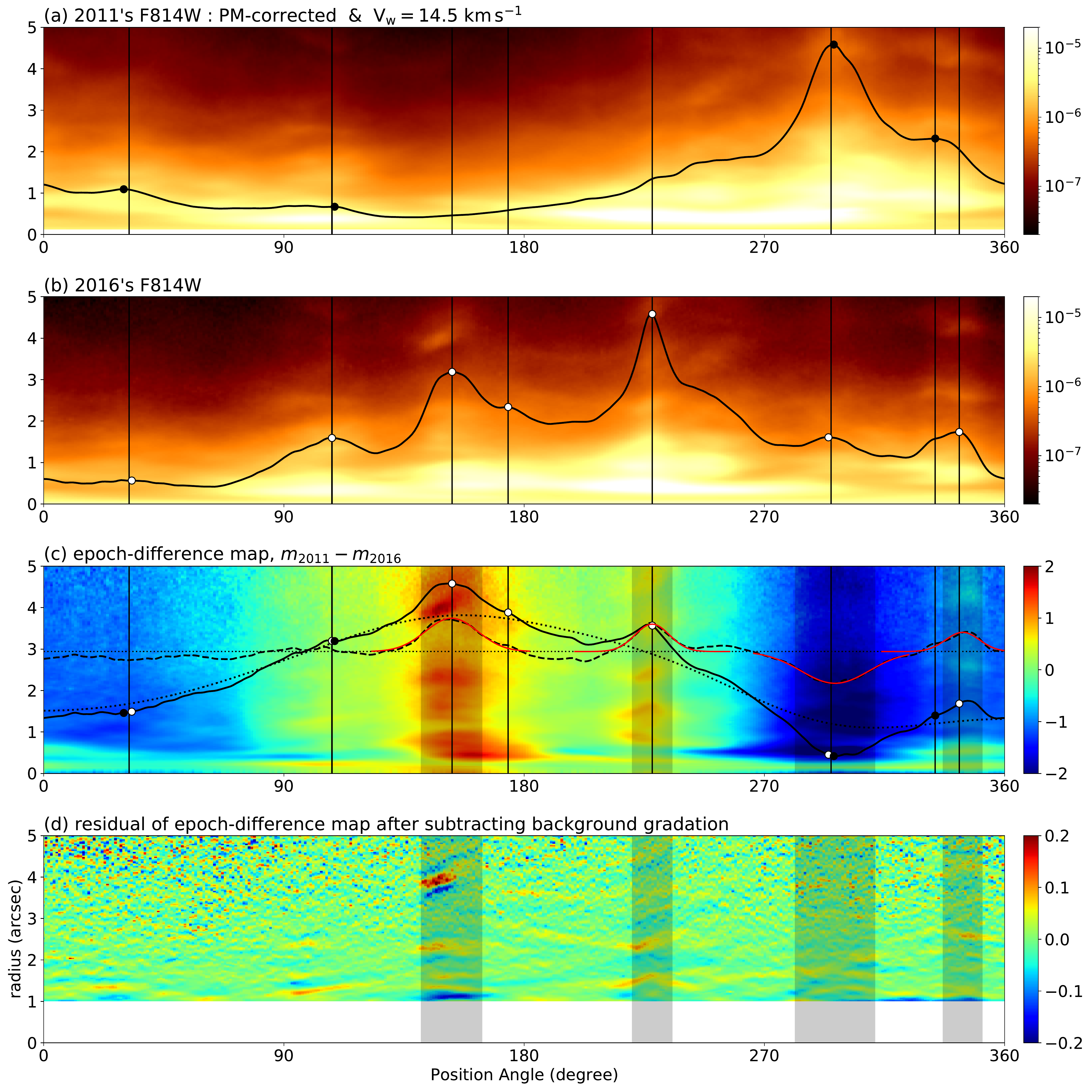}
  \caption{\label{fig:rtd}%
    (a) 2011's F814W image, presented in the radius versus position angle (PA)
    diagram. In order to directly compare to the corresponding 2016 image, its
    p.m.\ correction and the wind expansion for 4.95 years are already applied.
    The overlaid black curve shows the shape of brightness profile averaged
    in the radius range of 2\arcsec--5\arcsec, displayed along the PA.
    Four peaks (filled circles) are identified in this profile, indicating
    the PAs of searchlight beams in 2011 (see also Figure\,\ref{fig:bem}).
    (b) 2016's F814W image, with the shape of brightness profile averaged
    over radius (black solid curve) and its peaks (white filled circles).
    Units are in Jansky.
    (c) Magnitude difference map between 2011 and 2016 epochs (color coded)
    and the shape of its profile averaged over radius (black solid curve).
    The black and white filled circles on this profile are marked at the
    same PAs as in (a) and (b), respectively. Their PAs are indicated by
    eight black solid vertical lines (at 32\arcdeg, 108\arcdeg,
    153\arcdeg, 174\arcdeg, 228\arcdeg, 295\arcdeg, 334\arcdeg,
    and 343\arcdeg), and copied to (a) and (b) panels. The horizontal dotted
    line corresponds to the zero level in the epoch-difference profile.
    The background gradation in the epoch-difference map (dotted curve)
    is subtracted from the epoch-difference profile (black solid curve),
    resulting in the residual profile (dashed curve).
    The four significant (positive/negative) peaks in the residual profile are
    modeled to Gaussian functions (red solid curves) in order to determine the
    FWHM of searchlight beams. Gray-colored regions indicate the regions
    within the FWHMs of four significant beams.
    (d) The residual image of the epoch-difference map after subtracting
    the background gradation and ignoring the complex innermost region
    ($r<1\arcsec$; see Figure\,\ref{fig:clr}).
  }
\end{figure*}

Radius-angle diagrams are useful to verify the conical shapes of searchlight beams, as they present nearly vertical stretches as shown in Figure\,\ref{fig:rtd}(a)--(b).
We have first plotted the brightness profile averaged over radius ($>2\arcsec$) for each image and then its local peaks (black and white filled circles in 2011 and 2016 images, respectively) are identified as the PAs of the searchlight beams. The identified beams in the 2011 image are located at PAs of 30\arcdeg, 109\arcdeg, 296\arcdeg, and 334\arcdeg, while the beams in the 2016 image are identified at 33\arcdeg, 108\arcdeg, 153\arcdeg, 174\arcdeg, 228\arcdeg, 294\arcdeg, and 343\arcdeg.

Also in the magnitude difference map between the two epochs, the brightness profile averaged over radius is plotted (black solid curve) and it is further subtracted by its smoothed function with a Gaussian filter (black dotted curve) in order to ignore the background gradation (see also Section \ref{sec:cse}). The resulting profile (black dashed curve) represents the true amplitude of magnitude difference of the searchlight beams. The remaining searchlight beams with the amplitude greater than 0.3 mag are modeled by Gaussian functions (red solid curves): the beams at PAs of 153\arcdeg, 228\arcdeg, 296\arcdeg, and 344\arcdeg\ have the FWHMs of 12\arcdeg, 8\arcdeg, 15\arcdeg, and 7\arcdeg, respectively. The PA ranges within their FWHMs about the center PAs are shaded in gray in Figure\,\ref{fig:rtd}(c)--(d).

The remaining of these major bright regions along nearly straight lines in the radius-angle diagram for magnitude difference between 2011 and 2016 (Figure\,\ref{fig:rtd}(c)) indicates that the searchlight beams have significantly different relative brightnesses depending on the epochs. The positive values (red and yellow colors) in $\rm m_{2011}-m_{2016}$ along the southern beams indicate the relative excess in 2016, and the negative values (blue and cyan colors) remaining along the northwestern and northern beams indicate the relative excess in 2011.

\begin{figure*} 
  \epsscale{1.17}
  \plotone{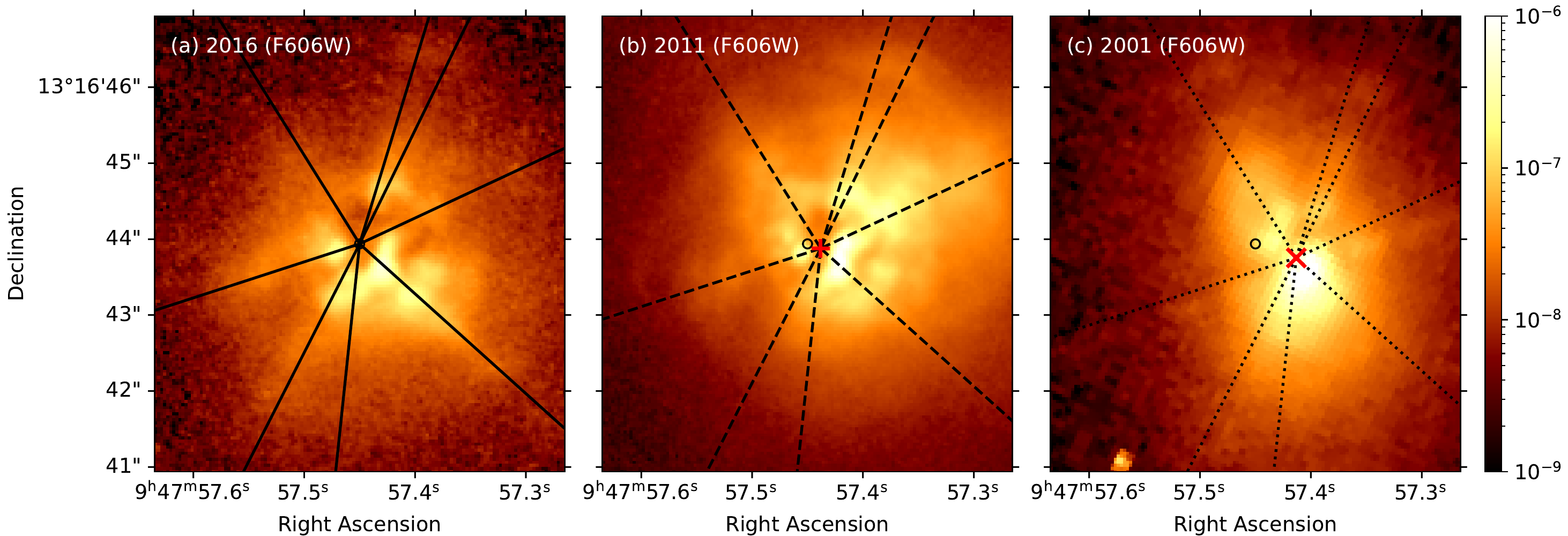}
  \caption{\label{fig:bem}%
    Searchlight beams inscribed on the F606W images taken in (a) 2016,
    (b) 2011, and (c) 2001. Eight straight lines indicate the locations of
    searchlight beams identified in two F814W images taken in 2011 and 2016;
    their PAs are 32\arcdeg, 108\arcdeg, 153\arcdeg, 174\arcdeg, 228\arcdeg,
    295\arcdeg, 334\arcdeg, and 343\arcdeg\ (see Figure\,\ref{fig:rtd} for
    detail). These beams are radially stretched from the stellar position
    in the 2016 image. Dashed and dotted lines in (b) and (c) are the same
    as the solid lines in (a) but the centers of the beams are accordingly
    shifted to the stellar positions at the corresponding epochs (red plus
    symbol in the 2011 image, and red cross symbol in the 2001 image).
    Open circle symbol in each image represents the stellar position at
    the 2016 epoch. In 2011, the southern beams (at PAs of 153\arcdeg\ and
    174\arcdeg) are absent, and the southwest beam (at PA of 228\arcdeg)
    is vague. In 2001, only two northern beams (at PAs of 32\arcdeg\ and
    334\arcdeg) are evident. Units are in Jansky.
    Note that the 2001 WFPC2 image has its original pixel 2.5 times larger
    than the pixel sizes in the other two images and its observation depth
    is much shallower than the others.
  }
\end{figure*}

The original images in 2016, 2011, and 2001 epochs with the F606W filter are compared in Figure\,\ref{fig:bem} with eight auxiliary lines representing the searchlight beams identified in Figure\,\ref{fig:rtd}(a)--(b) as the peaks of the brightness profiles integrated over radius in the F814W images taken in 2016 or 2011.
The very vertical shapes in the radius-angle diagrams indicate that the center of the beams is at the coordinate center (i.e., the stellar position). The eight auxiliary lines in Figure\,\ref{fig:bem}(a) drawn from the stellar position (red plus symbol) indeed well trace the beams.

The dashed and dotted lines overlaid in the 2011 and 2001 images, respectively, are the same as the solid lines in the 2016 images but the origin of the lines are shifted to the p.m.-corrected stellar positions at the corresponding epochs. The 2011 image does not exhibit bright patterns along the two southern lines (PA of 153\arcdeg\ and 174\arcdeg) and the southwestern beam (PA of 228\arcdeg), if exists, is extremely weak. The other five beams are apparent. The 2001 image additionally misses the eastern (PA of 108\arcdeg) and western (PA of 295\arcdeg) beams and the northern beam (PA of 343\arcdeg) is also unclear. We note that the bright stretch immediate south of the western 295\arcdeg-line is an artifact due to the diffraction spikes of saturated pixel. Although the observational depth is considerably shallower in the 2001 epoch, it is evident that the relative brightnesses of searchlight beams has changed along time.

Interestingly, the central optical nebula (size of $\sim2\arcsec$--$3\arcsec$) elongated along PA of $\sim$\,20\arcdeg--\,30\arcdeg\ has been reported in many papers based on imaging observations at epochs scattered over several decades and has been thought to be well modeled by an inclined bi-conical structure \citep[see also Figure\,\ref{fig:bem}(c);][]{bec69,dyc87,han98,ski98,mau00,lea06}. This bi-conical (or bipolar) nebula, however, is not apparent in the 2011 and 2016 {\it HST} images (see Figure\,\ref{fig:bem}(a)--(b))\footnote{A low-resolution image displayed in \cite{leb88} has an inverted triangle shape, which could be interpreted as similar to a lower-resolution version of the 2011 image shown in Figure\,\ref{fig:bem}.}, from which one may need to reassess the bipolar nebula model for IRC+10216. In the 2011 and 2016 images, the beam at the PA of $\sim$\,30\arcdeg\ is just one of a number of beams and, moreover, it is one of the faintest beam among the eight beams in the recent epochs.

We speculate that the searchlight beams are the stellar light illuminating the dust particles in the CSE, emerging through circumstellar holes near the star. If not, and if any motion of matter continuously ejected from the star is responsible for such structures, the persistence of their straight shapes with the observed lengths of 10\arcsec, regardless of the stellar p.m.\ during the 5 years, would require a speed of matter $>1000\,\kmps$. Moreover, the locations of the searchlight beams seem to be fixed (independent of the epochs) with respect to the stellar positions at the corresponding epochs.
The relative brightnesses of searchlight beams have changed likely in a rotating sense: the observed bright beams are spread around north in 2001 and 2011, while the bright beams in 2016 are located toward south--southeast. If it is true, one cycle of rotation of beam brightness change could be about 10 years. The reason for the rotation of relative brightness of the searchlight beams is unknown, but its timescale of 10-yr coincides with the timescale of long-term change found in the light curve using the photometric data from the Catalina Sky Survey in Appendix\,\ref{sec:lcr}. 
See also \citet{dyc91}, an earlier work that suggested a presence of 10-yr long term variation by inspecting near-infrared photometric and speckle interferometric observations. During the period from 1980 to 1990, the $K$-band flux and the point source contribution in it had increased, similar to the situation happening in the period dealt in this paper, albeit in a different wavelength.

\subsection{Circumstellar halo}\label{sec:cse}

\begin{figure*} 
  \epsscale{0.7}
  \plotone{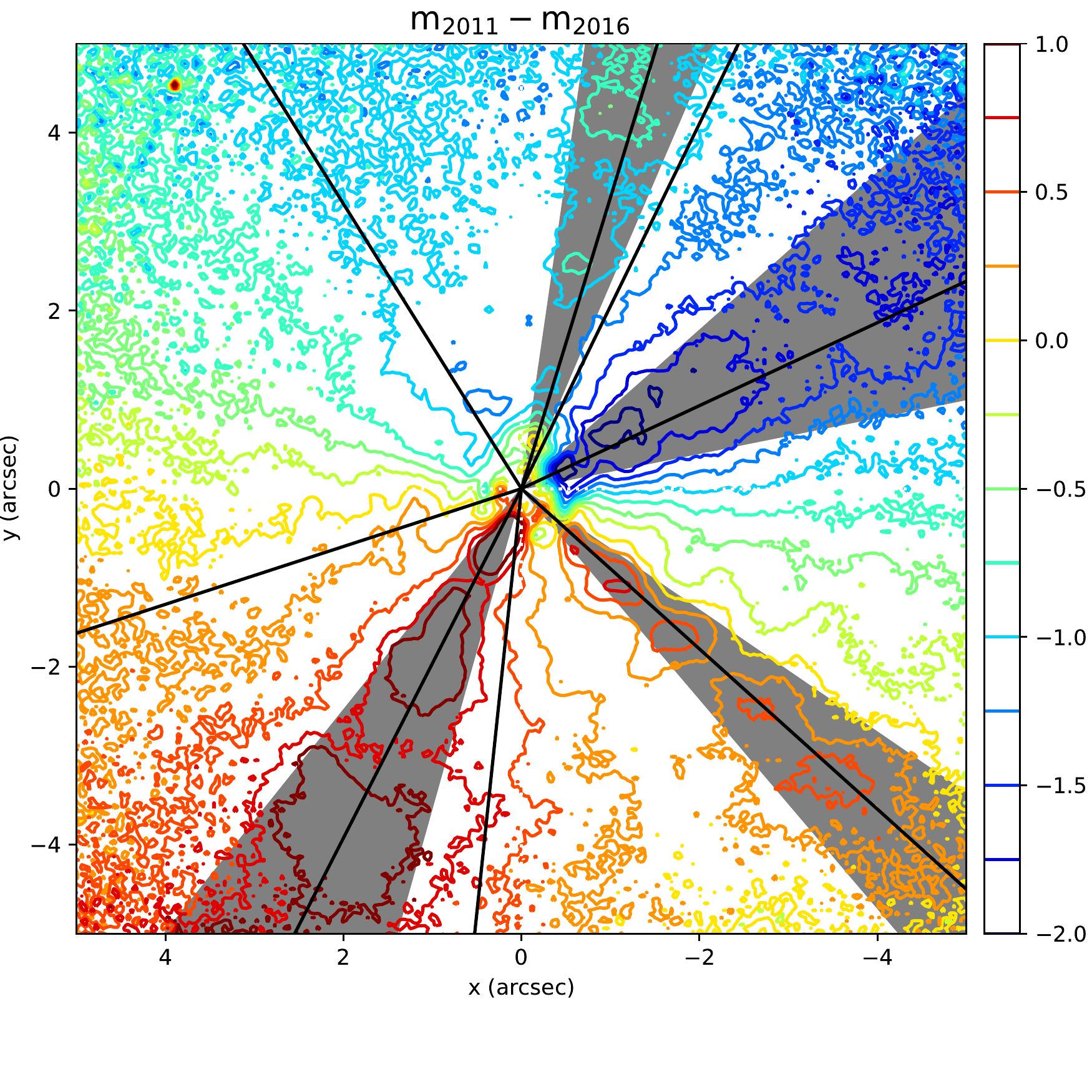}
  \caption{\label{fig:epd}%
    Same as Figure\,\ref{fig:cmp}(d) but in contours. Black straight
    lines indicate the position angles of searchlight beams identified
    in Section\,\ref{sec:bem}
    (see also Figures\,\ref{fig:rtd} and \ref{fig:bem}).
    Gray shaded regions indicate the four significant beam residuals
    within their FWHMs.
  }
\end{figure*}

As seen in the epoch-difference map eliminating the ring-like pattern (Figure\,\ref{fig:cmp}(d)), the change of the halo brightness from 2011 to 2016 has a positional dependence. In order to show the gradient of halo brightness change, the residual image is replotted in contours in Figure\,\ref{fig:epd}. A linear gradient is exhibited from $\rm m_{2011}-m_{2016}\la-2$ mag toward the northwest to $\ga+1$ mag toward the southeast, excluding the regions near PAs of 153\arcdeg, 228\arcdeg, 295\arcdeg, and 343\arcdeg\ where the searchlight beams are significantly remaining in the residual image.
The contour for the mean value $\sim-0.4$ mag of background gradation in the magnitude difference map between the two epochs is located nearly in the east-west direction (PA $\sim75\arcdeg$ and 75\arcdeg+180\arcdeg), making a nearly straight line.

\begin{figure*} 
  \epsscale{1.17}
  \plotone{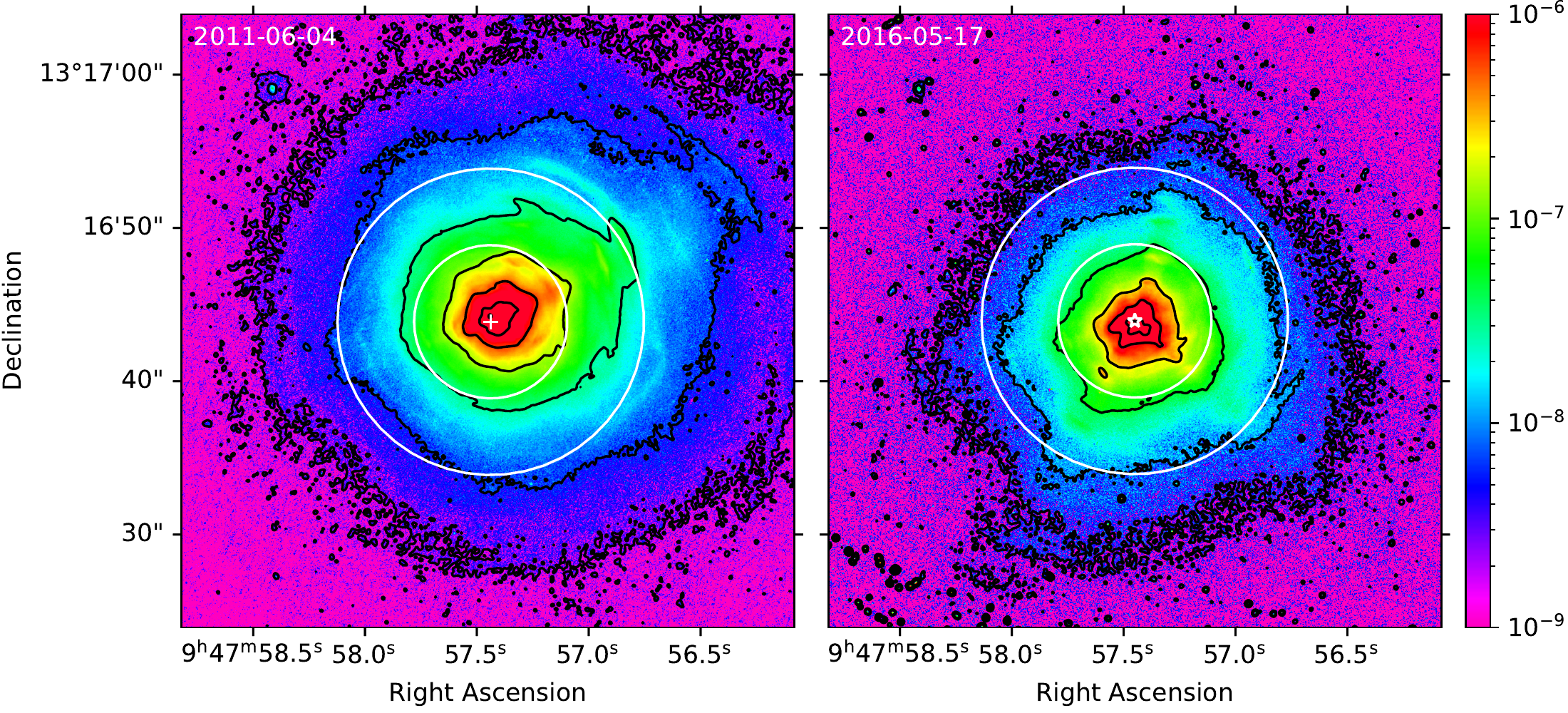}
  \caption{\label{fig:ext}%
    Extended halo of IRC+10216 in F814W at epochs of 2011 (left) and 2016
    (right). Units are in Jansky. The sky coordinates of the carbon star
    at the corresponding epochs are denoted by the plus and star symbols,
    respectively. Circles with radii of 5\arcsec\ and 10\arcsec\ from the
    stellar position at each epoch are overlaid for references.
    The contours are in levels of 1.5$\sigma$,
    4$\sigma$, 20$\sigma$, 100$\sigma$, 400$\sigma$, and 2000$\sigma$,
    where $\sigma$ indicates the noise level of each image: $1.5\times$ and
    $2.5\times10^{-9}$\,Jy/pixel in the left and right images, respectively.
  }
\end{figure*}

The halo brightness of images at the two epochs are compared in Figure\,\ref{fig:ext}, which exhibits the shifts of overall halo distributions slightly toward southern-southeastern direction in 2016 and significantly toward northwest direction in 2011. The length scale of halo variation indicated here, $r>10\arcsec$ (or $>1230$\,AU), is much larger than the length scale that can be explained by the wind dynamics ($\sim15$\,AU during 5 years) with the expansion speed of $\sim14\,\kmps$. This suggests that the halo brightness change is not due to moving material but more likely due to the illumination of stellar light. We speculate that the overall halo in the optical may be nothing but a searchlight beam {\it toward} us with the beaming angle that has been slightly changed from 2011 to 2016 epoch.

The overall halo brightness distribution in the 2016 image is centralized to the stellar position, likely implying that, in the above scenario, the postulated searchlight beam is nearly perpendicular to the plane of the sky. The concentrated light from the star toward us at this epoch may be the reason for the sudden appearance of the red compact source at the position of the star, as pointed out in Section\,\ref{sec:agb}.

\section{SUMMARY}\label{sec:sum}

This paper reports a careful examination of six imaging data toward IRC+10216 taken at three epochs using the {\it HST} with three filters. The six images are aligned using the positions of field stars and background galaxies, and the WCS coordinates of these images are assigned on the basis of the stellar positions and their proper motions available in {\it Gaia} Data Release 2. For comparison of images in the rest frame of the carbon star, the p.m.\ of the star based on radio continuum position monitoring is employed.

In the {\it HST}-F814W image in 2016, a red compact spot is spatially coincident with the radio p.m.-corrected stellar position.
In 2011, the reddest spot in the {\it HST} images is not at the same WCS but at the p.m.-corrected stellar position in the 2011 epoch. The reddest spot is either the carbon star itself or the dusty material surrounding the star, moving in the plane of the sky along with the star abiding by the stellar p.m.

Around the p.m.-corrected stellar positions, in both 2011 and 2016 epochs, the relatively red region is distributed in an $\Omega$-shape. The carved region in the $\Omega$-shape in the F606W--F814W color distribution map is coincident with the brightest regions in the optical images. It may suggest that the optically brightest central structures trace the light scattered by dusty matter slightly {\it above} the star along the line of sight. With this scenario in mind, the outer boundary of the $\Omega$-shaped distribution coincides with the dusty disk previously suggested by \citet{mur05} based on near-infrared polarization observations. 

By overlaying the ring-like patterns between two different epochs and taking into account the pattern expansion about the p.m.-corrected stellar position, the expansion velocity of the transverse wind components is measured for the first time. Making the p.m.\ correction is of particular importance as it affects the net outflow velocity distribution. The best-fit transverse velocity of the ring-like pattern, 12.5\,\kmps\,($d$/123\,pc), is compared to the line-of-sight velocity (14.5\,\kmps) measured from molecular line widths. The slightly smaller transverse wind velocity than the velocity perpendicular to the plane of the sky may support the elongated geometry of the CSE of IRC+10216 with the binary orbital plane located close to edge-on.

Seven searchlight beams are identified in the {\it HST} images taken in 2016. Among them, the northern beams were also found in 2011 and 2001 epochs at the same PAs about the corresponding p.m.-corrected stellar positions, but the southern beams were not clearly detected. Our best interpretation is that the searchlight beams are caused by the stellar light illumination through circumstellar holes near the star. The change of relative brightnesses among the identified searchlight beams, possibly in a rotating sense, tends to be in an about 10-year period. The reason for the change of their relative brightnesses is unknown, but this timescale is similar to a long-term variation in the light curve of IRC+10216.
The larger halo distribution over $>10\arcsec$-scale significantly changed from 2011 to 2016, which is also interpreted in the context of brightening of a postulated searchlight beam nearly pointing toward us.

\acknowledgments
We are grateful to the anonymous referee for the fruitful comments that helped to improve this paper.
This work is based on observations made with the NASA/ESA Hubble Space Telescope, obtained at the Space Telescope Science Institute (STScI), which is operated by the Association of Universities for Research in Astronomy, Inc., under NASA contract NAS5-26555. Support for program 14501 was provided by NASA through a grant from the STScI.
This work also has made use of data from the European Space Agency (ESA) mission {\it Gaia} (\url{https://www.cosmos.esa.int/gaia}), processed by the {\it Gaia} Data Processing and Analysis Consortium (DPAC, \url{https://www.cosmos.esa.int/web/gaia/dpac/consortium}). Funding for the DPAC has been provided by national institutions, in particular the institutions participating in the {\it Gaia} Multilateral Agreement.
HK acknowledges support by the National Research Foundation of Korea (NRF) grant funded by the Korea government (MIST) (No.\ 2021R1A2C1008928).
YO and YHC acknowledge the support by the Ministry of Science and Technology (MOST) of Taiwan through grants, MOST 109-2112-M-001-021- and MOST 109-2112-M-001-040, respectively.
PS is partly supported by the MOST of Taiwan under grant numbers MOST 104-2628-M-001-004-MY3 and MOST 107-2119-M-001-031-MY3, and by Academia Sinica under grant number AS-IA-106-M03.

\begin{deluxetable*}{cclccrccc}
  \tablecaption{\label{tab:list}%
    {\it HST} imaging datasets of IRC+10216 considered in this study. We list observation date, epoch, instrument/aperture, filter, pixel size (arcsec), exposure time (sec), proposal identification, primary investigator, and dataset.}
  \tablenum{1}

  \tablehead{
    \colhead{Obs.\ Date} & \colhead{Epoch}
    & \colhead{Instrument/Aperture}
    & \colhead{Filter} & \colhead{Pixel size} & \colhead{Exp.(s)}
    & \colhead{Prop.\ ID} & \colhead{PI} & \colhead{Dataset}}

  \startdata
  \tableline
  2001 Jan 07 & 2001.02 & WFPC2/WF3-FIX    & F606W & 0\farcs10 &  600\ \ \ \ &    8601 & P. Seitzer & U6722101R\\
  \tableline
  2011 Jun 04 & 2011.42 & WFC3/UVIS-CENTER & F606W & 0\farcs04 & 5407\ \ \ \ &   12205 &    T. Ueta & IBI901010\\
  -           & -       & WFC3/UVIS-CENTER & F814W & 0\farcs04 & 1974\ \ \ \ &       - &          - & IBI901020\\
  \tableline
  2016 May 17 & 2016.38 & WFC3/UVIS        & F606W & 0\farcs04 &  700\ \ \ \ &   14501 &     H. Kim & ID4403020\\
  -           & -       & WFC3/UVIS        & F814W & 0\farcs04 &  700\ \ \ \ &       - &          - & ID4403030\\
  -           & -       & WFC3/IR          & F098M & 0\farcs13 &   35\ \ \ \ &       - &          - & ID4403010\\
  \enddata

\end{deluxetable*}

\clearpage
\appendix
\section{Searchlight beams {\it versus} diffraction spikes}\label{sec:dfr}

\begin{figure*}[!h] 
  \epsscale{1.15}
  \plotone{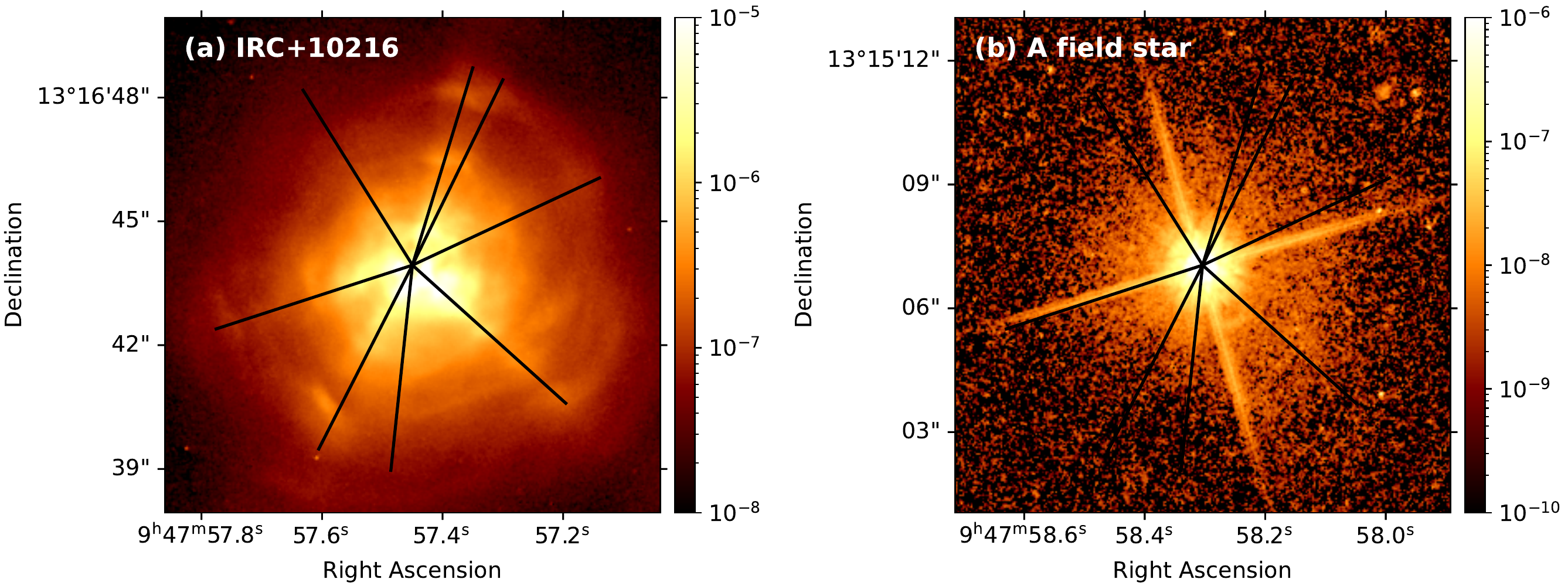}
  \caption{\label{fig:fld}%
    Comparison between the searchlight beams of IRC+10216 in the F814W
    image taken in 2016 and the diffraction spikes due to a saturated
    field star in the same image. Black lines mark (a) the position
    angles of searchlight beams of IRC+10216 and (b) their shift to
    the center of a field star. The searchlight beam at 108\arcdeg\
    is close to a diffraction spike in position angle, but the broad
    feature differentiates it from the diffraction effect.
  }
\end{figure*}

Figure\,\ref{fig:fld} confirms that the searchlight beams found in epoch 2016
are not instrumental artifacts caused by diffraction spikes.

\clearpage
\section{REVISITING LIGHT CURVE OF IRC+10216}\label{sec:lcr}

\citet{kim15a} presented the light curve provided by the Catalina Sky Survey experiment over the years 2005--2013 \citep{dra14}. The instrument does not use any filter in front of its wide bandwidth CCD detector. The effective wavelength is uncertain but is probably near 0.8--0.9\,$\mu$m and it may vary with phase because of stellar color variation. The Catalina data of IRC+10216 were fitted by a function, $m(t)=(A_0/2)\cos(2\pi(t-A_1)/A_2)+(A_3+A_4t)$, with constants $A_0$--$A_4$ as the fitting parameters. Over the period 2005--2013, it was found that the average flux regularly increased at a rate of $-0.16\pm0.004$\,mag\,yr$^{-1}$.

\begin{figure*}[!h] 
  \plotone{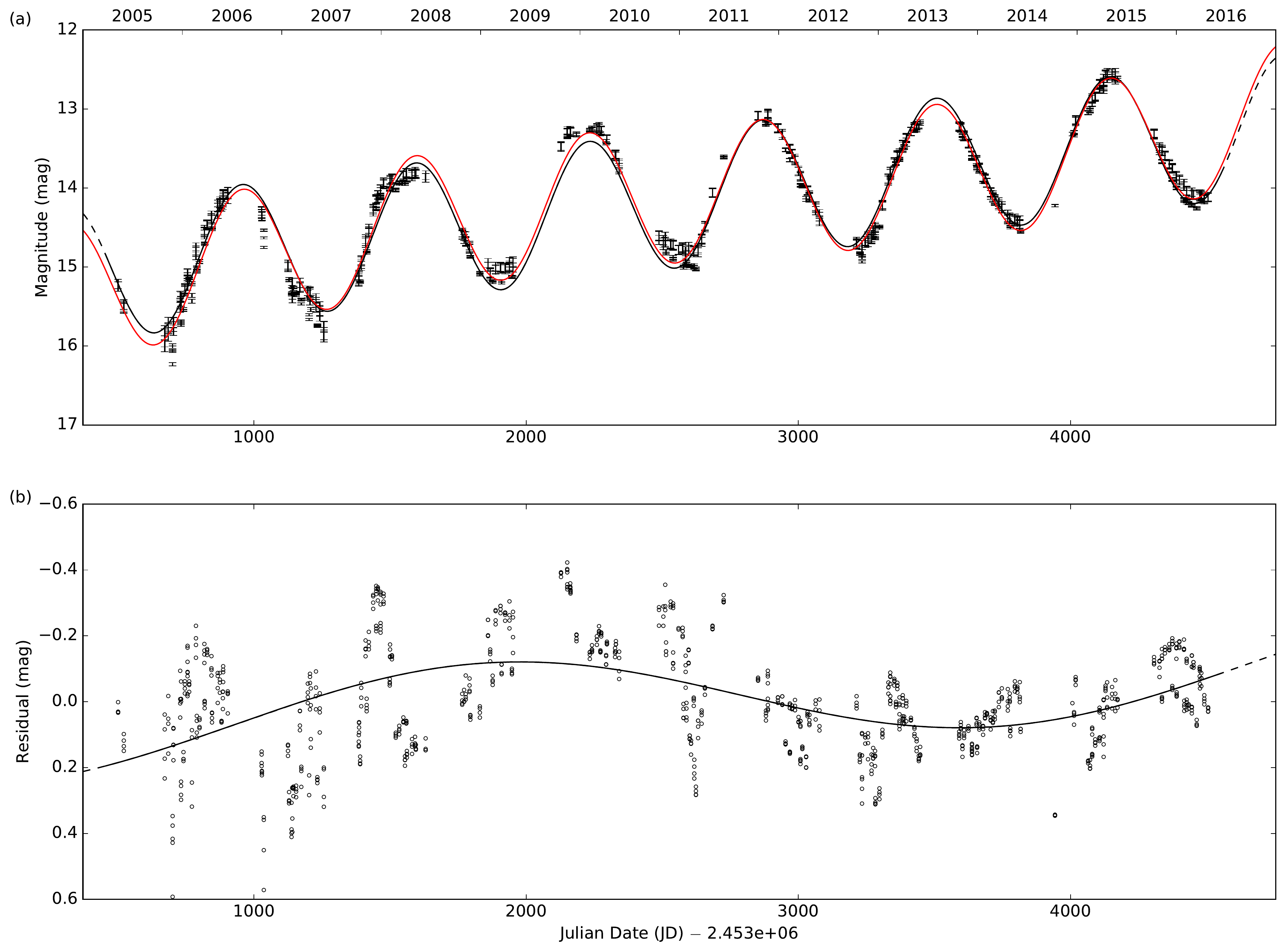}
  \caption{\label{fig:res}%
    (a) The light curve of IRC+10216 observed in the entire Catalina Sky
    Survey in the years of 2005 to 2016. The observed magnitude at epoch $t$
    is fit to a functional form composed of a sinusoidal term and a linear
    term, $m(t)=({\rm A_0}/2)\,\cos(2\pi\ t/{\rm A_2})+{\rm A_3+A_4}t$,
    where $t$ is date in JD $-$ 2453000 days.
    The fitting results are displayed by black curve.
    Red curve includes a correction for a long-term variation (see (b)).
    (b) The residual of the model fitting in (a) follows a sinusoidal
    function with a period of $\sim$\,10\,yr, which can be interpreted
    as a long-term variation of the light curve. 
  }
\end{figure*}

Here we update the model fitting including new photometric data observed in 2014--2016. In addition, we fix the unnecessary parameter $A_1$ to be $2.453\times10^6$ Julian Date.
The scientifically meaningful parameters obtained from fitting the entire monitoring of IRC+10216 in the Catalina Sky Survey in 2005--2016 are the amplitude of the sinusoidal variations $A_0=1.740\pm0.015$\,mag, the period of the sinusoidal variations $A_2=637\pm0.2$\,day, the brightness $A_3=15.236\pm0.012$\,mag at $2.453\times10^6$ Julian Date, and the brightening rate $A_4=-0.156\pm0.002$\,mag\,yr$^{-1}$ (see Figure\,\ref{fig:res}(a)). Therefore, the best fit including new epochs does not show considerable change in the fitting parameters.

We further analyze the light curve by inspecting the residual of the above fit from the Catalina photometric data. As seen in Figure\,\ref{fig:res}(b), the residual reveals sinusoidal variation with a period of $10.17\pm0.094$\,yr. The red curve in Figure\,\ref{fig:res}(a) demonstrates the light curve model including both 637-day and 10-yr sinusoidal variations with a brightening at a rate of $-0.156$\,mag per year. Note that the improvement on the deviation of the residuals of the fit is marginal, from 0.16 mag (black) to 0.14 mag (red), but the notable improvement is the fact that the final discrepancy does not show any sinusoidal trend.

\end{document}